\documentclass[twocolumn,english,showpacs,10pt]{revtex4}

\usepackage{babel,amsmath,amssymb,dcolumn}
\usepackage[dvips]{graphics}
\usepackage{graphicx}
\usepackage{epsfig}
\usepackage{setspace}[2]
\begin{document}

\title{ Inflationary Cosmology in RS-I}

\author{Michele Ferraz Figueir\'o}
\email{michele@fma.if.usp.br}

\affiliation{Instituto de F\'{\i}sica, Universidade de S\~{a}o Paulo \\
C.P. 66318, 05315-970, S\~{a}o Paulo-SP, Brazil}









\begin{abstract}
\centerline{\bf Abstract}
In this work, I intend to show a possible candidate of inflaton
potential $V(\phi)$ in a scenario of a brane world defined by a pair of branes (RS-I).  
\end{abstract}

\maketitle




The {\it Inflationary Cosmology} describes a phase in which our Universe evolves
through accelerated expansion in a short time period at high energy
scales. During this phase, our Universe is dominated by a
potential $V(\phi )$ generated by a homogeneous scalar field $\phi (t)$
called {\it inflaton}. This potential must obey {\it slow-roll
  conditions} $\{ \epsilon , | \eta | \ll 1\}$ where $\epsilon$ and $\eta$
are the {\it slow-roll parameters}. These parameters are given by {\cite{cosmo}}
\begin{equation}
\epsilon(\phi ) = \frac{M^2_{PL}}{2}\left ( \frac{V'}{V} \right )^2 
\end{equation}
and
\begin{equation}
\eta (\phi) = M^2_{PL}\frac{V''}{V} \, .
\end{equation}





We can calculate the spectral index $n(\phi)$ and its derivate for
this potential $V(\phi )$  

\begin{equation}
n - 1 = -6\epsilon + 2 \eta 
\end{equation}
and 
\begin{equation}
\frac{dn}{dlnk} = - 16 \epsilon \eta + 24 \epsilon ^2 + 2\zeta 
\end{equation}
where 
\begin{equation}
\zeta = M^4_{PL}\frac{V'V'''}{V^2} \, .
\end{equation}

The amount of inflation that occurs is described by the {\it number of e-foldings
  N}, is given by
\begin{equation}
N \equiv ln \frac{a(t_{end})}{a(t)} \equiv \int^{t_{end}}_{t} H dt
\approx \frac{1}{M^2_{PL}} \int^{\phi}_{\phi _{end}} \frac{V}{V'}d\phi ,
\end{equation}
where $\phi _{end}$ is defined by $\epsilon(\phi _{end})= 1$ if
inflation ends through violation of the slow-roll conditions.

The {\it Brane Cosmology} describes cosmological models with extra
dimensions. A lot of interest in brane cosmology arose with a publication of two
papers by {\it Randall} and {\it Sundrum} in the 90s. They propose a new higher-dimensional mechanism for
solving the hierarchy problem building two models, {\it RS-I} \cite{brane1} and {\it
  RS-II} \cite{brane2}. In these two models, they consider that the
Standard Model particles and forces,
with exception of gravity are confined to a {\it four-dimensional subspace},
within the {\it five-dimensional spacetime (bulk)}, referred to as {\it 3-brane}.    
Many researches have been done around this new cosmology as e.g.
{\cite{bra4}}. We choose the {\it inflation in branes} \cite{infbra1},
\cite{bra1}, \cite{bra2}, \cite{bra3}.

In this study, we consider that inflation might arise from the
interaction potential between a 3-brane and anti-3-brane which are
parallel and widely separated in five-dimensional Anti de Sitter space
($AdS_5$). The background is identical to that considered in the RS-I model {\cite{brane1}}.

The potential between the branes is given by \cite{infbra1}

\begin{equation}
V(\phi ) \sim M^4_{5D} \xi(\phi / M_{5D})(1-e^{(-|\phi|/m)}).
\end{equation}





\begin{figure*}[htb!]
\begin{center}
\epsfig{file=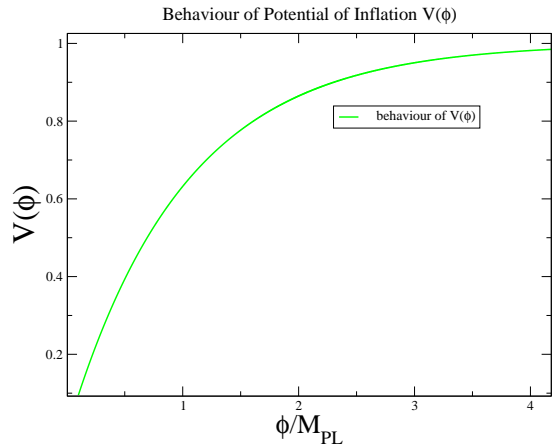,scale=0.3,angle=-90}
\caption{\small Behaviour of the potential $V(\phi )$ as a function
  $\phi$.}
\label{fag03}
\end{center}
\end{figure*}


\begin{figure*}[htb!]
\begin{center}
\epsfig{file=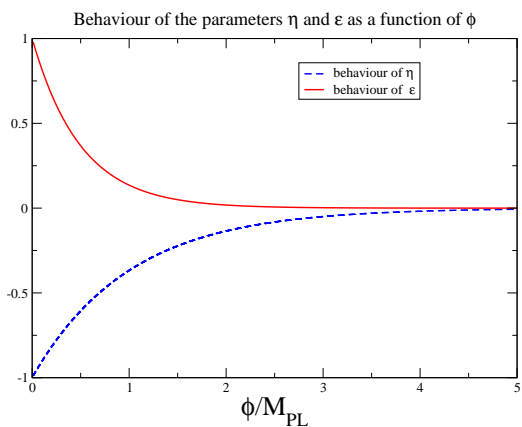,scale=0.3,angle=-90}
\caption{\small Behaviour of the slow-roll parameters $\epsilon$ and
  $\eta$ as a function of $\phi$.}
\label{fag03}
\end{center}
\end{figure*}




The figures 1 and 2 show us the behaviours of the potential
$V(\phi)$ and of the slow-roll parameters as a function of $\phi$.





The spectral index $n$ and its derivate $dn/dlnk$ can be related to
$N$, respectively, as

\begin{equation}
\frac{1}{2}(n-1) = -\frac{1}{N},
\end{equation}
and
\begin{equation}
\frac{1}{2}\frac{dn}{dlnk}= - \frac{1}{N^2}.
\end{equation}

Setting $N=70$ (as usually done in inflationary scenarios) leads to

\begin{equation}
n \approx 0,9714 
\end{equation}
and
\begin{equation}
\frac{dn}{dlnk} \approx -0,0004,
\end{equation}
in excellent agreement with observational data from WMAP ($0.94 \leq n
\leq 1.00$ and $-0.02 \leq dn/dlnk \leq 0.02$, \cite{infbra2})



\section{Acknowledgement}
This work was supported by CNPq.


\end{document}